\pdfoutput=1                              

\documentclass[10pt,a4paper]{article}     

\usepackage[T1]{fontenc}
\usepackage[utf8]{inputenc}
\usepackage{lmodern}                      

\usepackage{amsmath,amssymb,amsfonts}
\usepackage{graphicx}                     

\usepackage{cite}                         
\usepackage[hidelinks]{hyperref}          

\newcommand{\bse}{\begin{subequations}}
\newcommand{\ese}{\end{subequations}}

\usepackage{hyperref}
\hypersetup{
    colorlinks=true,   
    linkcolor=red,     
    citecolor=blue,    
    urlcolor=blue      
}

\DeclareUnicodeCharacter{200B}{}
 
\begin{document}
\title{\bf Schwinger effect in dynamical holographic QCD with a supercritical region}

\date{}
\maketitle
\vspace*{-0.3cm}
\begin{center}
{\bf Leila Shahkarami$^{1}$, Farid Charmchi$^{2}$}\\
\vspace*{0.3cm}
{\it {School of Physics, Damghan University, Damghan 36716-45667, Iran
}} \\
\vspace*{0.3cm}
{\it  {${}^1$l.shahkarami@du.ac.ir}, {${}^2$farid.charmchi@gmail.com} }
\end{center}

\begin{abstract}
We study the phase structure of QCD matter using a dynamical Einstein--Maxwell--Dilaton holographic model, using both thermodynamic and dynamical observables. Depending on the warp factor, the model admits either a standard confinement/deconfinement transition or a first-order \textit{specious confinement}/deconfinement transition ending at a critical end point (CEP), giving rise to a rich phase diagram with a supercritical region. 
We probe this structure using both thermodynamic (heat capacity) and dynamical (squared speed of sound, IR wall) observables. We show that the loci of maxima in the heat capacity and minima in the sound speed define two distinct crossover lines emanating from the CEP and extending into the supercritical region, each tracing a different separation between confined-like and deconfined-like matter. As a dynamical probe, the persistence of the IR wall introduces another separation line, not emanating from the CEP, and reveals a triangular region, before the CEP, where the system is thermodynamically confined but dynamically deconfined. 
We further study the Schwinger effect as a nonperturbative probe of vacuum instability, determining the critical and threshold electric fields in both confined and deconfined phases. In the specious confined phase and in the confined-like phase beyond the critical end point, these fields depend on temperature and chemical potential, unlike in the standard confined phase, and we are able to trace their behavior for the first time including the supercritical region.
Our results highlight the complementarity of thermodynamic and dynamical probes in mapping the QCD phase diagram and, in particular, establish the Schwinger threshold and critical fields as sensitive diagnostics of confinement not only in the known phase transitions but also in the supercritical regime.


\end{abstract}

Keywords: Schwinger effect; Supercritical phase; Confinement; Dynamical holographic QCD.
\section{Introduction}
Understanding the phase structure of strongly interacting matter, as governed by Quantum Chromodynamics (QCD), remains one of the central challenges in modern high-energy and nuclear physics. In the low-energy regime, where the coupling is strong, QCD exhibits confinement—quarks and gluons are bound within hadrons. At sufficiently high temperatures or baryon densities, however, QCD undergoes a transition to a deconfined phase known as the quark-gluon plasma (QGP). Mapping the nature of this transition and exploring the behavior of physical observables near the transition, and in particular near the possible critical end point(s), are of both theoretical and experimental importance.
Establishing whether such critical points exist, and identifying characteristic signatures of its vicinity, remains one of the central open questions in the study of strongly interacting matter.

The concept of \emph{supercritical behavior}, originally developed in condensed matter physics for classical fluids~\cite{widom} and later extended to electronic fluids~\cite{superelectronic}, has recently attracted attention in strongly interacting matter. In systems with a first-order transition line ending at a critical end point, the supercritical region beyond the (critical end point) CEP no longer exhibits a sharp phase boundary—the two phases become indistinguishable. Nevertheless, various \emph{crossover lines} can be defined, such as the Widom line (thermodynamic origin, traced by maxima of susceptibilities)~\cite{widom}, the Fisher–Widom line (structural)~\cite{fisherwidom}, and the Frenkel line (dynamical)~\cite{frenkel}. These concepts have been generalized to QCD-like theories~\cite{widomQCD,superholoBaggioli}, where different probes of supercriticality provide signatures of the proximity to the CEP. The universality of these supercritical crossovers makes them particularly powerful: even when the CEP is not directly accessible, one can locate it indirectly by identifying where distinct supercritical separation lines converge.

The non-perturbative nature of QCD at low energies presents significant challenges for most analytical methods. However, substantial progress has been made in exploring the QCD phase diagram at vanishing or small chemical potential using lattice QCD simulations. These studies demonstrate that at low temperatures, QCD matter exists in the form of hadrons due to quark confinement. As temperature is raised, a transition to the deconfined phase of quark and gluon plasma takes place. Lattice studies have firmly established the crossover nature of the confinement–deconfinement transition at zero chemical potential, along with an accompanying chiral symmetry restoration~\cite{ref1,ref2}. The critical temperature for this transition is frequently cited around 150–175 MeV.

Despite ongoing efforts to develop methods for extrapolating to nonzero chemical potentials, lattice QCD faces the well-known sign problem, which renders it less effective in regimes of high baryon density. In these high-density regions, a first-order transition line and a possible critical end point (CEP) are conjectured to exist, supported by various models and functional calculations~\cite{crit1,crit2,crit3,crit4,crit5,crit6,crit7,crit8,crit9}. 
The search for this critical end point is a primary objective of experimental programs such as RHIC's Beam Energy Scan.
Since lattice extrapolation methods are typically reliable only near zero density and lose precision as density increases, the region where the first-order transition line would terminate and beyond the CEP, where the supercritical crossover would start, remains mostly speculative in lattice studies.

In this context, the gauge/gravity duality—and in general, holography—has emerged as a powerful framework for studying strongly coupled gauge theories in regimes inaccessible to perturbative techniques or lattice simulations~\cite{Maldacena2,Maldacena3,Maldacena1,Solana}. Originally formulated to relate type IIB string theory on AdS$_5 \times S^5$ to $\mathcal{N}=4$ super-Yang-Mills theory~\cite{Maldacena1}, holography has since been extended to QCD-like theories through both top-down and bottom-up constructions. While top-down models derive from string theory and capture important qualitative features of QCD matter and its phase structure, they often include unwanted conformal symmetry and extra degrees of freedom, which do not directly match QCD \cite{topdownp1,topdownp2}. Phenomenological bottom-up hard-wall and soft-wall models of QCD are limited in that their background geometries are fixed and do not solve Einstein’s equations. 
Moreover, they may suffer from problems with the area law of the Wilson loop and chiral condensate \cite{sofwallp1,sofwallp2,sofwallp3}. A significant common limitation of both top-down and many bottom-up approaches is that their confined phases are independent of temperature, making it difficult to study the thermodynamics of confinement without including extremely challenging $1/N$ corrections. This motivates the introduction of Einstein-Maxwell-Dilaton (EMD) models, which address these issues directly.

Among bottom-up models, EMD models~\cite{emd1,emd2,emd3} stand out. These models, which provide solutions to the five-dimensional gravity action equations, incorporate free parameters tuned to reproduce desirable properties of real QCD. Crucially, they allow for a dynamical treatment of background geometries, thereby incorporating temperature and chemical potential dependence. This enables EMD models to realize rich thermodynamic behavior, confinement, and chiral symmetry breaking in a unified framework.

A particularly interesting phenomenological EMD model was proposed by Dudal and Mahapatra~\cite{emdDMspecious}. They introduced two distinct scale factor profiles: one with a quadratic form generating a Hawking–Page phase transition, which translates to a standard confinement/deconfinement transition; and another with a logarithmic form yielding a first-order small/large black hole transition, corresponding to what the authors termed a \emph{specious confinement}/deconfinement transition. In the first case, the confined phase is described by a thermal-AdS background, while in the second case, the confined phase exists at finite temperature, supported by a small black hole horizon in the bulk. The latter construction directly addresses the problem of temperature independence in most of confined holographic phases. This feature provides a fertile ground not only for exploring confinement at finite temperature and density, but also for studying rich structures such as the QCD critical end point and even supercritical phenomena.

In the first model with a standard confined phase, the string tension is, as expected, independent of temperature and chemical potential in the confined phase. Dudal and Mahapatra argued that the string tension in the specious confined phase is finite and nearly temperature-independent below $T_c$, remains non-zero at $T=T_c$, and drops discontinuously at the transition point. This behavior is qualitatively consistent with quenched lattice QCD observations\footnote{Lattice studies show that in full QCD with dynamical quarks, the physical string tension—a direct order parameter for confinement—vanishes at $T_c$, while in quenched QCD, below the critical temperature, it decreases only mildly with temperature and remains finite at $T_c$, consistent with a first-order transition, before vanishing above it~\cite{stlattice1,stlattice2,stlattice3}. While this provides valuable information about confinement near the transition, the sign problem limits further exploration at high densities.} \cite{stlattice1,stlattice2,stlattice3}, providing a crucial validation point for this phenomenological EMD model, despite its simplified nature. Above $T_c$, the linear part of the interquark potential disappears in both models, indicating a fully deconfined phase, in agreement with the dissolution of confinement. 
Since the phase diagram of the second model contains a first-order transition line ending at a critical end point, it is possible to trace thermodynamic and dynamical probes into the supercritical region. In this work, we exploit this rich structure to study the Schwinger effect in different phases, while also exploring the potential of the model to capture supercritical crossovers through dynamical observables.

The Schwinger effect, a nonperturbative phenomenon involving the creation of charged particle-antiparticle pairs from vacuum in the presence of a strong external electric field, is significant from both theoretical and experimental standpoints. It can serve as a dynamical probe of vacuum instability and confinement in strongly coupled systems. Originally formulated in Quantum Electrodynamics (QED)~\cite{Schwinger}, the Schwinger effect has been widely studied in field theory~\cite{manton1,periodic,assist,scatter1,critical,semi1,kinetic2,kinetic1,exact1,Schaxioninflation} and holography, where two complementary approaches are used: (i) potential analysis using the Nambu–Goto action or Wilson loops \cite{semenoff,potential,Sch1,Sch2,Sch3,confin1,confin2,Sch5,Sch6,Sch7,confinrev,dehghani,us2,me,Sch8}, and (ii) the imaginary part of the Dirac-Born-Infeld (DBI) action on probe branes with external fields \cite{decay1,decay2,equen,magneticdecay,LF,Sch8}. Both methods provide insight into vacuum decay and the thresholds for particle production in different phases.

A central outcome of holographic studies is the identification of two critical electric fields. One of them called $E_c$, above which pair production is unsuppressed, was first predicted by field theoretical investigations under weak-field conditions \cite{manton1}, where its existence remained unconfirmed due to being outside the calculable range and only was confirmed by holographic studies \cite{semenoff} which are, by contrast, well-suited for strong coupling and intense fields.
The other one, called $E_s$, is characteristic of confinement, below which even massless pairs cannot be produced. In deconfined phases, $E_s$ typically vanishes, while in confining phases it remains finite, making the Schwinger effect a sensitive dynamical probe of confinement.

While extensive research has been conducted on the holographic Schwinger effect using diverse holographic models, further lessons remain to be gleaned from this non-perturbative phenomenon.
Motivated by this, in this paper, we revisit this diagnostic in the framework of the Dudal–Mahapatra EMD model \cite{emdDMspecious}, focusing on the distinction between the standard confined phase and the specious confinement background. 
While in the standard confined phase (in the first model), both the IR wall position and $E_s$​ are independent of temperature and chemical potential, the explicit temperature dependence of the confined phase in the latter case enables, for the first time, a detailed study of how the IR wall position and $E_s$ vary with both temperature and chemical potential. 
An interesting observation arises when examining the theory at low temperature and nonzero chemical potential: the IR wall and $E_s$​ at small T differ from their counterparts at zero chemical potential. This suggests that finite density continues to influence the infrared dynamics even as $T\to 0$, possibly reflecting an inherent breaking of Lorentz symmetry in the strongly coupled plasma at finite density.
We show that $E_s$ behaves as a sharp dynamical order parameter for confinement, consistent with quenched lattice results and with the discontinuous behavior of the string tension.

Furthermore, the model’s rich phase structure allows us to explore how these dynamical probes extend into the supercritical region, providing a novel perspective on the interplay between confinement, criticality, and dynamical instabilities.
We systematically analyze the emergence of separation lines in the supercritical region.
By examining the loci of extrema in the specific heat and the squared speed of sound as thermodynamic and dynamical response functions, respectively, we identify distinct crossover lines that continue smoothly from the CEP into the supercritical domain. 
Remarkably, although both originate from the CEP, the heat-capacity line terminates sharply at finite chemical potential, whereas the sound-speed line extends smoothly to higher densities and vanishes only on the $T=0$ axis. 
We further employ the persistence of the IR wall as a dynamical criterion for confinement, which delineates a separation line not attached to the CEP but instead branching from the first-order transition line. 
This distinction is consistent with the behavior of dynamical crossovers in classical fluids. 
The combination of these probes results in a refined supercritical phase diagram containing multiple separation lines, with each carrying complementary information about the nature of QCD matter beyond the critical point.

The paper is organized as follows. In Section~2 we review the holographic EMD model of interest. Section~3 presents the thermodynamic structure and phase diagram. We also study some quantities to define possible lines separating the phases beyond the critical end point. Section~4 is devoted to the potential analysis in the presence of a static external electric field, including the behavior of the critical fields $E_c$ and $E_s$. We moreover study a dynamical separation line to complete the proposal of a revised phase diagram of this model, including the supercritical region.
 In Section~5, we summarize our findings and discuss possible extensions.

\section{Einstein--Maxwell--Dilaton background}

Einstein--Maxwell--Dilaton (EMD) models have been extensively employed to explore the QCD phase structure and the properties of strongly coupled QCD matter. By appropriately tuning the scale factor and/or the dilaton potential, one can construct black hole solutions that reproduce the expected features of confinement in the dual field theory. This flexibility allows for a richer phase diagram, including scenarios where the confined phase admits a well-defined notion of temperature, enabling the study of temperature dependence for various observables below the critical temperature $T_c$.

Among many holographic approaches to real QCD, EMD models stand out as phenomenological bottom-up constructions that are self-consistent solutions of the Einstein equations. They have been widely investigated to extract thermodynamic, transport, and spectral properties of QCD matter and to clarify the details of its phase structure. Here, we briefly describe the specific EMD setup introduced and analyzed in \cite{emdDMspecious}, which possesses the features required for our purposes. In particular, one choice of scale factor leads to black hole solutions even in the confined phase, thereby introducing a notion of temperature on the field theory side and yielding a more realistic phase structure compared to most holographic models used to study QCD and its phase diagram.

We start from the five-dimensional EMD action containing a $\mathrm{U}(1)$ gauge field $A_M$, a dilaton field $\phi$, and the metric $g_{MN}$:
\begin{align}\label{action}
S_\textrm{EMD}=-\frac{1}{16\pi G_5}\int\! d^5\!x\sqrt{-g}
\left[{\cal R}-\frac{h(\phi)}{4}F_{MN}F^{MN}-\frac{1}{2}\partial_M\phi\partial^M\phi-V(\phi)
\right].
\end{align}
Here, $G_5$ is the five-dimensional Newton constant, $g$ is the determinant of the metric, $F_{MN}$ is the field strength tensor of the gauge field, $h(\phi)$ is the kinetic gauge function, and $V(\phi)$ is the dilaton potential. The time component of the gauge field is taken to have a nonzero boundary value, corresponding to a finite baryon number density in the dual field theory.

We adopt the following ansatz for the metric, gauge field, and dilaton field:
\begin{align}\label{metric}
  ds_e^2&=(g_e)_{M N} dX^M dX^{N}={\cal H}_e(z)\left(-G(z) dt^2+\frac{dz^2}{G(z)}+d\vec{x}^2\right),\nonumber\\
A_M&=A_t(z), ~~ \phi=\phi(z).
\end{align}
Here $G(z)$ is the blackening function, which breaks Lorentz symmetry and ${\cal H}_e(z)=\frac{R^2 e^{2{\cal A}_e(z)}}{z^2}$,
where $R$ is the AdS radius and ${\cal A}_e(z)$ is a scale factor describing a deformation of AdS space. The radial coordinate $z$ runs from the AdS boundary at $z=0$ to the interior.

Instead of fixing $h(\phi)$ and $V(\phi)$ \emph{a priori}, as in \cite{emdDMspecious}, we follow the \emph{potential reconstruction method} (see e.g., \cite{reconst1,reconst2,reconst3,reconst4,reconst5,hajilou}), in which ${\cal A}_e(z)$ and $h(z)$ are chosen to reproduce known QCD features. For example, the function $h(z)$ can be determined by requiring the meson spectrum in the dual field theory to lie on a linear Regge trajectory.

With the metric ansatz in \eqref{metric} and the choice of $h(z)=e^{-cz^2-{\cal A}_e(z)}$,
the equations of motion for the three fields in the action can be solved to obtain the holographic dual of QCD matter:
\begin{align}\label{solution}
G(z)&=1-\frac{1}{\int^{z_h}_0\! dx\!~x^3\!~e^{-3{\cal A}_e(x)}}\int^{z}_0\! dx\!~x^3\!~e^{-3{\cal A}_e(x)}\nonumber\\
&+\frac{2c\mu^2}{(1-e^{-cz_h^2})\int^{z_h}_0\! dx\!~x^3\!~e^{-3{\cal A}_e(x)}}
\begin{vmatrix}
\int^{z_h}_0\! dx\!~x^3\!~e^{-3{\cal A}_e(x)}&\int^{z_h}_0\! dx\!~x^3\!~e^{-3{\cal A}_e(x)}\\[0.15cm]
\int^{z}_{z_h}\! dx\!~x^3\!~e^{-3{\cal A}_e(x)}&\int^{z}_{z_h}\! dx\!~x^3\!~e^{-3{\cal A}_e(x)}
\end{vmatrix},\nonumber\\
\phi'(z)&=\sqrt{6\left({{\cal A}_e^{\prime}}^{2}-{\cal A}_e^{\prime\prime}-2{\cal A}_e^{\prime}/z\right)},\nonumber\\
A_t(z)&=\mu\frac{e^{-cz^2}-e^{-cz_h^2}}{1-e^{-cz_h^2}},\nonumber\\
V(z)&=-\frac{3z^2Ge^{-2{\cal A}_e}}{R^2}\left[{\cal A}_e^{\prime\prime}+{\cal A}_e^{\prime}\left(3{\cal A}_e^{\prime}-\frac{6}{z}+\frac{3G'}{2G}\right)-\frac{1}{z}\left(
-\frac{4}{z}+\frac{3G'}{2G}\right)\right].
\end{align}
This solution is obtained by imposing the boundary conditions $G(z_h)=0$ and $\lim_{z \to 0} G(z)=1$, where $z_h$ is the black hole horizon. The parameter $\mu$ is the chemical potential of the boundary theory. The solution is analytic in terms of the single scale factor ${\cal A}_e(z)$, whose form determines the geometry and hence the nature of the phase transitions between gravitational solutions, and their interpretation in the boundary theory.

In what follows, we work with two specific forms of ${\cal A}_e(z)$ chosen in \cite{emdDMspecious}:
\begin{align}\label{Ae1}
{\cal A}_e^{(1)}(z)&=-\bar{a} z^2,\\\label{Ae2}
{\cal A}_e^{(2)}(z)&=-\frac{3}{4}\ln(bz^2+1)+\frac{1}{2}\ln(dz^3+1)-\frac{3}{4}(fz^4+1).
\end{align}
Each has one or more free parameters. The model parameters, including $c$ and those appearing in the scale factors, are fixed by matching holographic predictions to lattice QCD results at zero chemical potential. For instance, comparing the meson mass spectrum to heavy meson states yields $c=1.16~\mathrm{GeV}^2$. Matching the confinement--deconfinement transition temperature at $\mu=0$ to $T_c \approx 270~\mathrm{MeV}$ determines the parameters of the quadratic and logarithmic scale factors as
\begin{align}\label{params}
 \bar{a}=\frac{c}{8},
\end{align}
and
\begin{align}\label{params2}
 b=\frac{c}{9},\quad d=\frac{5c}{16},\quad f=b.
\end{align}

As will be shown in the next section, the quadratic scale factor leads to a first-order Hawking--Page transition between thermal AdS and a black hole phase, corresponding to the standard confinement/deconfinement transition. The logarithmic scale factor, by contrast, yields a small-to-large black hole transition, which has been interpreted as a specious confinement/deconfinement transition in \cite{emdDMspecious}. The small black hole phase is termed specious confined because, although it retains some confining features---such as a linear potential at large distances at low temperatures---it also exhibits a small but nonzero Polyakov loop. Interestingly, the behavior of the speed of sound and entropy density as functions of temperature in this phase closely resembles that seen in unquenched lattice results, especially near $T_c$, suggesting a possible connection between the specious confined phase and genuine confinement.

\section{Thermodynamic quantities}

We now compare the thermodynamics of the two EMD models with quadratic and logarithmic scale factors.
For a general choice of ${\cal A}_e(z)$, the Hawking temperature and Bekenstein--Hawking entropy of the black hole solutions are given by
\begin{align}\label{T}
T&=\frac{z_h^3\!~e^{-3{\cal A}_e(z_h)}}{4\pi\int^{z_h}_0\! dx\!~x^3\!~e^{-3{\cal A}_e(x)}}
\left[
1+\frac{2c\mu^2\left(e^{-cz_h^2}\int^{z_h}_0\! dx\!~x^3\!~e^{-3{\cal A}_e(x)}
-\int^{z_h}_0\! dx\!~x^3\!~e^{-3{\cal A}_e(x)}e^{-cx^2}\right)}{\left(1-e^{-cz_h^2}\right)^2}
\right],\\\label{s}
S_{ _\textrm{BH}}&=\frac{L^3e^{3{\cal A}_e(z_h)}}{4G_5z_h^3}.
\end{align}
The free energy is obtained by integrating the entropy in the grand canonical ensemble at fixed chemical potential,
\begin{align}\label{free}
 F=\int_{z_h}^{\infty} S_{ _\textrm{BH}}(z_h)T'(z_h) dz_h.
\end{align}
We use the condition that the free energy of thermal-AdS is zero. 

\subsection{Quadratic scale factor}

\begin{figure}[h]
\begin{center}
\includegraphics[width=6.8cm]{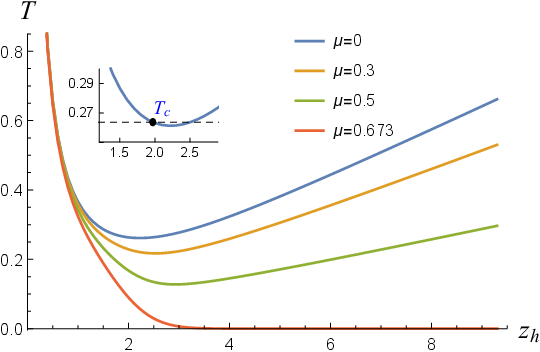}\hspace{.3cm}
\includegraphics[width=6.8cm]{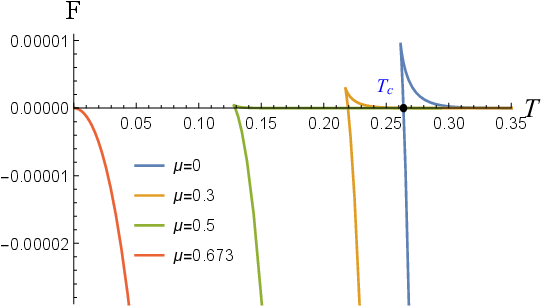}
\end{center}
\caption{\footnotesize 
Left: Temperature versus $z_h$ for various values of the chemical potential.
Right: Free energy as a function of temperature for the same values of $\mu$.}
\label{temp1}
\end{figure}

We begin with the model defined by the quadratic scale factor \eqref{Ae1}. Notice that all the results in the figures of this section and the subsequent one are in units GeV. 
The left panel of Fig.\,\ref{temp1} shows the temperature $T(z_h)$ for several values of the chemical potential $\mu$. Two distinct branches appear:  
(i) a \emph{stable} branch at small $z_h$, where $T$ decreases with increasing $z_h$, i.e., decreasing the size of the black hole; and  
(ii) an \emph{unstable} branch at larger $z_h$, where $T$ increases with $z_h$, contrary to the behavior of a physical black hole.

The stability analysis of the black hole solutions is confirmed by the free energy curves in the right panel of Fig.\,\ref{temp1}. For lower values of the chemical potential, such as $\mu=0$ (blue curve), the temperature profile exhibits a minimum $T_{\mathrm{min}}$ below which no black hole solution exists; in this regime the thermal AdS solution, with zero free energy, is favored. For $T_{\mathrm{min}} < T < T_c$, the free energy of the black hole branch is positive, indicating that thermal AdS remains the physical solution. Only for $T > T_c$ does the large black hole, small $z_h$, branch become dominant, as its free energy becomes negative. This signals a first-order Hawking--Page transition from thermal AdS to a large AdS black hole while increasing the temperature. $T_c = 0.264\,\mathrm{GeV}$ in the zero chemical potential case.

\begin{figure}[h]
\begin{center}
\includegraphics[width=6.8cm]{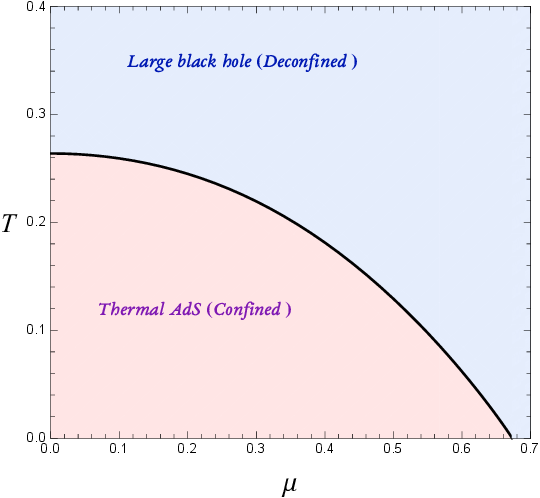}
\end{center}
\caption{\footnotesize 
The transition temperature versus $\mu$ for the model with the quadratic scale factor.}
\label{tcrit1}
\end{figure}

Turning on the chemical potential $\mu$ shifts the curves: at a given $T$ in the stable branch, charged black holes have smaller $z_h$ than uncharged ones, i.e., it has a larger size with respect to the uncharged ones, leading to a monotonic decrease of $T_c(\mu)$, in qualitative agreement with lattice QCD. This is shown in Fig.\,\ref{tcrit1}. 
Increasing $\mu$ reduces the extent of the unstable branch until it disappears at $\mu_c = 0.673\,\mathrm{GeV}$, where $T_c$ approaches zero. Below $T_c$ the system is in the thermal AdS (confined) phase, while above $T_c$ it is in the large black hole (deconfined) phase. The resulting $(T,\mu)$ phase diagram thus contains a line of first-order transitions separating confined and deconfined phases. As discussed in \cite{emdDMspecious}, the behavior of $q\bar{q}$ free energy supports this phase assignment.

\subsection{Logarithmic scale factor}

\begin{figure}[h]
\begin{center}
\includegraphics[width=6.8cm]{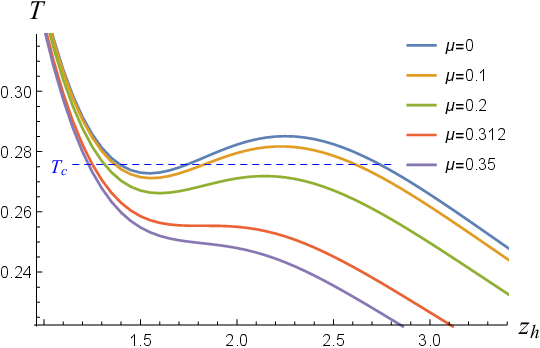}\hspace{.3cm}
\includegraphics[width=6.8cm]{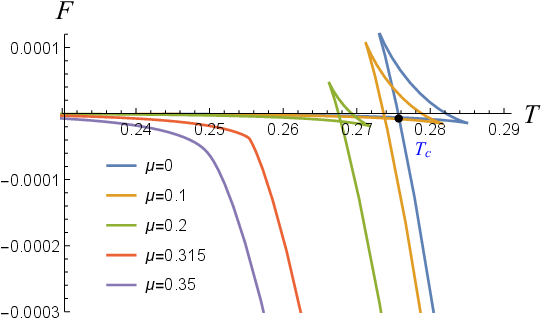}
\end{center}
\caption{\footnotesize 
Left: Temperature versus $z_h$ for various values of the chemical potential.
Right: Free energy as a function of temperature for the same values of $\mu$.}
\label{temp2}
\end{figure}

We now turn to the EMD model with a logarithmic scale factor \eqref{Ae2}.  
The temperature $T(z_h)$ and free energy $F(T)$ are displayed in Fig.\,\ref{temp2}. 
Here, three branches are present: a stable large black hole branch (negative slope), an intermediate unstable branch (positive slope), and a metastable small black hole branch (negative slope). At small $\mu$, the free energy displays a swallowtail structure, characteristic of a first-order transition between the small and large black hole phases. The two tails corresponding to small and large black hole solutions have smaller (negative) free energy than the other parts (the base of the swallowtail). The tip of the swallowtail marks the transition temperature which is $T_c = 0.276\,\mathrm{GeV}$ for $\mu=0$.

\begin{figure}[h]
\begin{center}
\includegraphics[width=6.8cm]{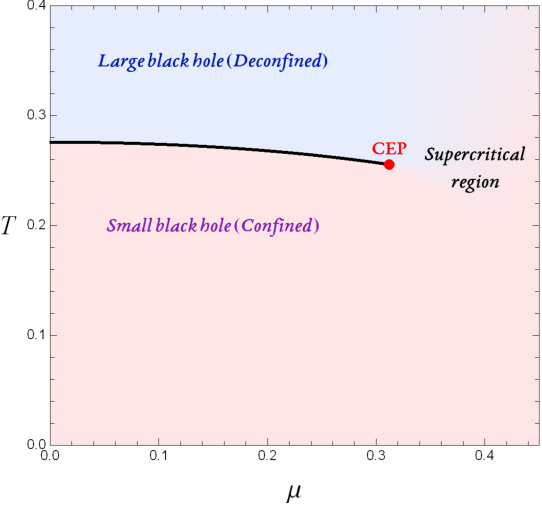}
\end{center}
\caption{\footnotesize 
Phase diagram $T-\mu$ for the model with the logarithmic scale factor. Solid line shows the first-order transition temperature to the deconfined phase and the red dot is the CEP.}
\label{tcrit2}
\end{figure}

As $\mu$ increases, $T_c$ decreases and the unstable branch shrinks, vanishing at a critical point $\mu_{ _{\mathrm{CEP}}} = 0.312\,\mathrm{GeV}$. 
At this point, the transition becomes second order, analogous to a liquid--gas critical endpoint. The phase diagram in Fig.\,\ref{tcrit2} summarizes these results: at low $\mu$, a solid line separates the small black hole phase from the large black hole phase, corresponding respectively to confined and deconfined phases in the dual theory \cite{60DM,61DM}. This line terminates at a second order critical end point (CEP).

The interpretation of the small black hole phase as a true confined phase has limitations, as discussed in \cite{emdDMspecious}. Nonetheless, several observables, including the entropy and speed of sound, show good qualitative agreement with unquenched lattice QCD near $T_c$, suggesting that the specious confinement picture may capture relevant features of strongly coupled matter (see \cite{emdDMspecious}).

Above CEP, i.e.\ for $\mu>\mu_{ _{\mathrm{CEP}}}$, there exists a region analogous to the so-called supercritical phase in classical fluids. 
An important question is whether any separation line between confined-like and deconfined-like regimes can be identified in this region by means of thermodynamic or dynamical probes. 
To this end, we consider two key observables that are highly sensitive to the phase transition.

\subsubsection*{-- Specific heat and the Widom line}

\begin{figure}[ht]
\begin{center}
\includegraphics[width=6.8cm]{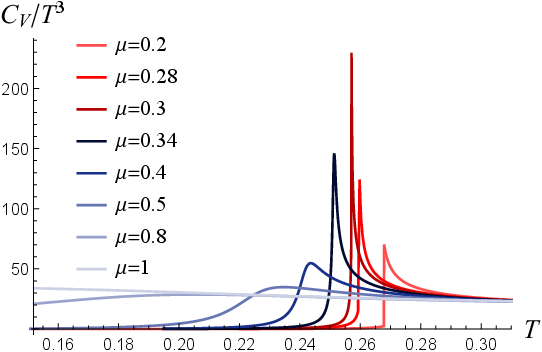}
\end{center}
\caption{\footnotesize 
Heat capacity rescaled by $T^3$ as a function of temperature for different values of the chemical potential, in the second model. Red (blue) lines show the results for chemical potentials below (above) CEP.}
\label{cv}
\end{figure}

The specific heat, related to the derivative of entropy with respect to temperature as
\begin{equation}\label{eqcv}
C_V = T \left( \frac{\partial S_{{}_\textrm{BH}}}{\partial T} \right)_{\mu},
\end{equation}
encodes valuable information about phase transitions. Figure~\ref{cv} shows the rescaled specific heat, $C_V/T^3$, as a function of temperature for several values of the chemical potential. 

For $\mu<\mu_{ _{\mathrm{CEP}}}$, displayed by the red curves, the specific heat shows delta-function-like spikes at the first-order transition temperature, as expected.
For $\mu>\mu_{ _{\mathrm{CEP}}}$ (blue curves), corresponding to the supercritical region, the sharp spike disappears and instead a smooth maximum appears. 
This maximum is broad and mild far from the CEP, but becomes sharper, narrower, and of larger magnitude as the CEP is approached. 
Such universal behavior is typical of thermodynamic response functions in systems with a first-order transition line terminating at a CEP, independent of microscopic details, and reflects the proximity to the critical point. 

\begin{figure}[h]
\begin{center}
\includegraphics[width=6.8cm]{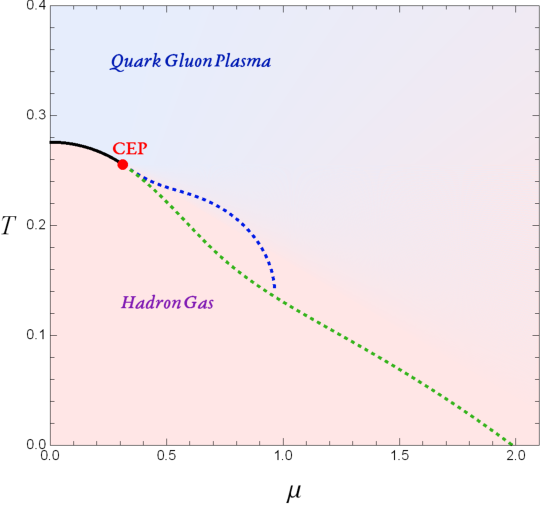}
\end{center}
\caption{\footnotesize 
Temperature $T$ versus chemical potential $\mu$ phase diagram of the model with logarithmic scale factor. The separation lines given by the extrema of heat capacity and squared sound speed are depicted by blue and green lines, respectively.}
\label{phasediagram1}
\end{figure}

The locus of these maxima defines a thermodynamic separation line known as the \emph{Widom line}. 
This line, depicted by the dotted blue curve in the phase diagram (Fig.~\ref{phasediagram1}), separates confined-like from deconfined-like behavior in the supercritical region.

\subsubsection*{-- Speed of sound and dynamical separation}

A complementary dynamical observable sensitive to phase transitions is the speed of sound. 
In the grand canonical ensemble, the squared speed of sound is given by
\begin{equation}\label{eqcs2}
C_s^2 = \frac{S_{{}_\textrm{BH}}}{T(\frac{\partial S_{{}_\textrm{BH}}}{\partial T})_\mu
+\mu\left(\frac{\partial n}{\partial T}\right)_\mu},
\end{equation}
where, using the holographic dictionary, the charge density $n$ can be obtained from the near-boundary expansion of the bulk gauge field $A_t$. Comparing the relation of $A_t$ given in Eq.\eqref{solution} with the near-boundary expansion $A_t(z)=\mu-n z^2+\ldots$, we arrive at
\begin{equation}\label{density}
n=\frac{\mu c}{e^{cz_h^2}-1}.
\end{equation}

\begin{figure}[h]
\begin{center}
\includegraphics[width=6.8cm]{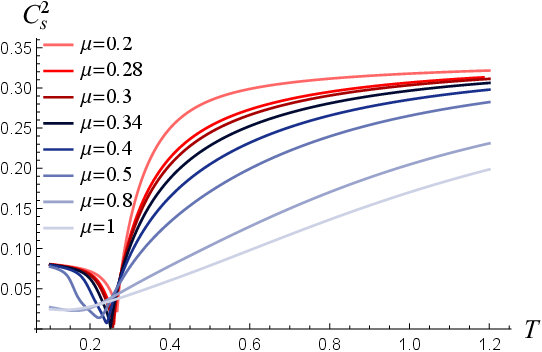}
\end{center}
\caption{\footnotesize 
Squared sound speed as a function of temperature for different values of the chemical potential, in the second model. Red (blue) lines show the results for chemical potentials below (above) CEP.}
\label{cs2}
\end{figure}

Figure~\ref{cs2} displays $C_s^2(T)$ for various values of the chemical potential. 
In all cases, both below and above $\mu_{ _{\mathrm{CEP}}}$, the sound speed approaches $1/3$ at very high temperatures, consistent with the conformal invariance in this limit. 
For $\mu<\mu_{ _{\mathrm{CEP}}}$, $C_s^2$ exhibits a sharp dip near the deconfinement transition temperature, in qualitative agreement with minima seen in unquenched lattice QCD results \cite{105DM,106DM}, although in lattice QCD they are smooth.

In the supercritical region, the minimum of $C_s^2(T)$ behaves analogously to the specific heat: it is broad and shallow far from the CEP, but becomes sharper and deeper as the CEP is approached. 
The position of this minimum at fixed $\mu$ coincides with the inflection point of the corresponding $T(z_h)$ curve. 
Tracing these minima yields a dynamical separation line beyond the CEP, shown as a dotted green curve in Fig.~\ref{phasediagram1}. 
This line continues smoothly from the CEP to lower temperatures, eventually terminating at the point where the transition temperature is reduced to zero. 
Beyond this point, the system remains in the deconfined phase at all temperatures. 

The depth of the minimum in $C_s^2(T)$ decreases as $\mu \to \mu_{ _{\mathrm{CEP}}}$, approaching zero at the critical point. 
For $\mu>\mu_{ _{\mathrm{CEP}}}$, the minimum increases again to positive values with increasing chemical potential. The lowest minimum value occurs at the CEP.

It is important to emphasize that, while both separation lines originate exactly at the critical end point (CEP), they do not coincide beyond that point: the locus obtained from the specific-heat maxima (the thermodynamic/Widom-like line) lies at higher temperatures for a given 
$\mu$ and then terminates sharply—dropping to zero temperature—at a relatively low value of $\mu$; by contrast, the locus defined by the minima of the squared speed of sound 
$C_s^2(T)$ (a dynamical separation line) continues smoothly toward larger $\mu$ and reaches the temperature axis at a much higher chemical potential. In other words, both lines stem from the CEP but follow distinct trajectories in the supercritical region, reflecting different physical diagnostics.

\section{Potential analysis in the presence of a static electric field}

We now investigate how the EMD models introduced above respond to an external static electric field $E$. Our aim is to compute the total potential of a static quark--antiquark pair, separated by a distance $L$ and held fixed in space under the influence of $E$.

The potential is obtained from the expectation value of a rectangular Wilson loop of spatial width $L$ and infinite temporal extent. In the gauge/gravity duality, this quantity is dual to the on-shell Nambu--Goto action of an open string stretching from the quark and antiquark positions on the boundary into the bulk. 

The creation of infinitely heavy quark particles is severely suppressed. Therefore, to study the Schwinger effect, we must consider quarks of finite mass rather than infinitely heavy ones. Following the Semenoff and Zarembo's prescription \cite{semenoff}, this is achieved holographically by attaching the string endpoints to a probe D3-brane located at a finite radial position $z_0$ in the bulk, rather than at the boundary ($z=0$).

For the $q\bar{q}$ system, it is convenient to work in the \emph{string-frame} metric, obtained by transforming the Einstein-frame metric via a dilaton-dependent rescaling,
\begin{align}\label{metricstring}
 ds_s^2=e^{\sqrt{\!\frac{2}{3}}\phi}ds_e^2={\cal H}_s(z)\left(-G(z) dt^2+\frac{dz^2}{G(z)}+d\vec{x}^2\right),
\end{align}
where ${\cal H}_s(z)=\frac{R^2 e^{2{\cal A}_s(z)}}{z^2}$ and ${\cal A}_s (z)= {\cal A}_e(z) + \sqrt{\frac{1}{6}}\,\phi(z)$.

The Nambu--Goto action for the open string is
\begin{align}\label{actionNG}
 S_{\mathrm{NG}}=T_f\int d\tau d\sigma \sqrt{-\det \left[(g_s)_{ab}\right]},
\end{align}
where $T_f=\frac{1}{2 \pi \alpha'}$ is the fundamental string tension and $(g_s)_{ab}$ is the string-frame metric induced on the worldsheet. We choose the static gauge $\tau = t$ and $\sigma = x$, with $x$ one of the spatial boundary coordinates. The quark and antiquark are placed symmetrically at $x = \pm L/2$. Under these conditions, Eq.\,\eqref{actionNG} becomes
\begin{align}\label{actionNGon}
  S_{\mathrm{NG}}=T_f{\cal T}\int dx\; {\cal H}_s(z)\sqrt{G(z)+z'^2}.
\end{align}

Since the action does not explicitly depend on $z$, the corresponding effective Hamiltonian is conserved along the $z$ direction. Its value is fixed by the conditions $z(0) = z_c$ and $z'(0) = 0$, where $x=0$ denotes the tip of the U-shaped string profile and $z_c$ is its turning point. Integrating the Hamiltonian relation yields the separation length $L$ between the quarks 
\begin{align}\label{L}
 L(z_c)=2 \int_{z_0}^{z_c} dz\; \frac{\sqrt{{\cal G}(z,z_c)}}{\sqrt{{\cal F}(z)-{\cal F}(z_c)}},
\end{align}
where $z_0$ denotes the position of the probe D3-brane, ${\cal G}(z,z_c)=\frac{G(z_c){\cal H}_s(z_c)^2}{G(z)}$, and ${\cal F}(z)=G(z){\cal H}_s(z)^2$. Defining the dimensionless parameter $a\equiv z_c/z_0$ and the new coordinate $y\equiv z/z_c$, we arrive at
\begin{align}\label{La}
 L(a)=2 z_0 a\int_{1/a}^1 dy \;\frac{\sqrt{{\cal G}(y z_0 a,z_0 a)}}{\sqrt{{\cal F}(y z_0 a)-{\cal F}(z_0 a)}}.
\end{align}
Then, by adding the contribution from the potential associated with the electric field $E$ to the potential energy of the $q\bar{q}$, the total potential is obtained as
\begin{align}\label{v}
 V_{\mathrm{total}}(L)=2T_f R^2 z_0 a\int_{1/a}^1 dy \;\frac{{\cal H}_s(y z_0 a)^2\sqrt{G(y z_0 a)}}{\sqrt{{\cal F}(y z_0 a)-{\cal F}(z_0 a)}}-E L.
\end{align}
 In all the following results we set $T_f R^2=1$ and $R=1\,\mathrm{GeV}^{-1}$.

The system’s critical response to the electric field plays a central role in both the potential analysis and the phase structure study. Therefore, in the next section we determine the critical electric fields for both models before proceeding to present and discuss the resulting total potential.
\subsection{Infrared wall}

In the holographic description of a quark–antiquark pair, the \emph{infrared (IR) wall} is defined as the deepest radial position $z_{\mathrm{IR}}$ in the bulk that a connected U-shaped string can reach while remaining physical. In general, the bulk geometry contains a black hole horizon at $z=z_h$, except in the confined phase of the first model, where the physical background is thermal AdS and no horizon is present. Therefore, $z_{\mathrm{IR}}$ lies in the interval $0< z_{\mathrm{IR}}< z_h$ in the specious confined phase, and $0< z_{\mathrm{IR}}<\infty$ in the standard confined phase of the first model. In such cases, the allowed string embedding extends from the boundary (or the probe brane at $z=z_0$) into the bulk, but cannot cross the IR wall. The IR wall signals a dynamical limit to the space accessible to the string, and its presence is associated with the dual gauge theory being in a confining phase.

A practical way to determine $z_{\mathrm{IR}}$ is to identify the radial position where the function
\begin{align}\label{est}
\sigma_s(z) \;=\; \sqrt{-g_{tt}(z)\,g_{xx}(z)}\;=\;\sqrt{\mathcal{F}(z)}
\end{align}
in the string-frame metric attains its minimum. Here, $\sigma_s(z)$ is proportional to the square root of the determinant of the induced metric along the Wilson loop direction and is directly related to the effective string tension density. The IR wall is then given by
\begin{align}\label{zir}
z_{\mathrm{IR}} = \underset{z}{\mathrm{arg\,min}}\ \sqrt{\mathcal{F}(z)},
\end{align}
and a connected string can only extend down to $z_c \leq z_{\mathrm{IR}}$.  

Applying this analysis to our models, we find that an IR wall appears only in the confined phase, as expected from general holographic considerations. 
In the first model with the quadratic scale factor, one has $G(z)=1$ in the thermal AdS background corresponding to the confined phase. The effective string tension is then given by $\sigma_s(z)={\cal H}_s(z)$. For this choice of scale factor, the dilaton field admits an analytic expression,  
\begin{equation}
   \phi(z) = z\sqrt{3\bar{a}\left(3+2\bar{a}z^2\right)}+3\sqrt{\frac{3}{2}}\sinh^{-1}\left(\sqrt{\frac{2\bar{a}}{3}}z\right) ,
\end{equation}
so that the location of the IR wall can be determined analytically by solving  
\begin{equation}
   \frac{d\sigma_s(z)}{dz}=0,
\end{equation}
which yields $z_{\mathrm{IR}}=\frac{1}{\sqrt{2 \tilde{a}}}$. Importantly, this position is independent of both temperature and chemical potential.  

\begin{figure}[h]
\begin{center}
\includegraphics[width=6.8cm]{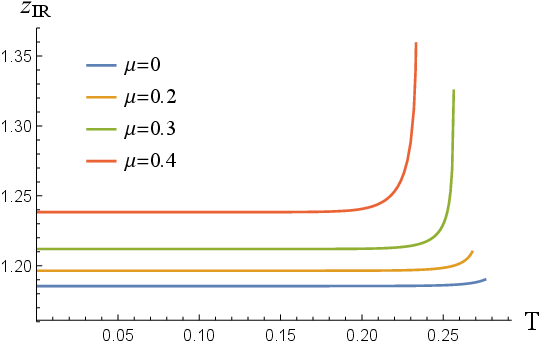}\hspace{.3cm}
\includegraphics[width=6.8cm]{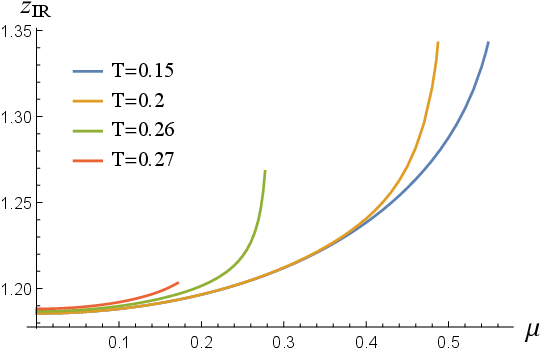}
\end{center}
\caption{\footnotesize 
IR wall position $z_{\mathrm{IR}}$ as a function of temperature $T$ (left graph) and chemical potential $\mu$ (right graph), for the model with logarithmic scale factor. }
\label{IR}
\end{figure}

In contrast, in the second model with the logarithmic scale factor, the confined phase is realized by the small black hole background, and the effective string tension is $\sigma_s(z)=\sqrt{G(z)}\,{\cal H}_s(z)$. In this case, the dilaton does not admit a closed analytic form, and consequently the IR wall location cannot be obtained analytically. Instead, $z_{\mathrm{IR}}$ must be determined numerically by analyzing the extremum of the effective string tension density. We find that in this model the IR wall position depends nontrivially on both temperature and chemical potential, as illustrated in Fig.\,\ref{IR}. 
This difference reflects the distinct ways the bulk geometries of the two models encode confinement and will have direct consequences for the critical electric fields discussed in the next subsection.

By analyzing the graphs depicted in Fig.\,\ref{IR}, several observations can be made.
First, for $\mu<\mu_{ _{\mathrm{CEP}}}$, the IR wall $z_{\mathrm{IR}}$ only exists in the confined phase in all cases.
It disappears abruptly at the confinement–deconfinement transition, where the bulk geometry switches to the deconfined large black hole solution and the accessible radial interval for the string extends continuously to $[0,z_h]$.
Importantly, $z_{\mathrm{IR}}$ does not merge smoothly with the horizon as the transition is approached.
Instead, it retains a finite value just before the transition point and then vanishes (disappears) discontinuously once the system enters the deconfined phase.

At small values of $T$ and/or $\mu$, the position of $z_{\mathrm{IR}}$ remains nearly constant, reflecting the stability of the confining background at low excitation.
However, as $T$ or $\mu$ increases, especially near the transition to the deconfined phase, $z_{\mathrm{IR}}$ begins to shift more noticeably.
In particular, enhancement of either temperature or chemical potential drives $z_{\mathrm{IR}}$ to larger values, meaning that the connected string can extend deeper into the bulk before turning.

Physically, this behavior reflects the progressive weakening of the confining force due to medium screening: thermal effects at finite $T$ and density effects at finite $\mu$ soften the flux tube between quark and antiquark.
Since the IR wall controls the string–breaking field $E_s​$, we expect this temperature and chemical potential dependence to directly influence the behavior of $E_s​$, as will be demonstrated in the next subsection.

As can be seen from the plot of the IR wall against temperature in Fig.\,\ref{IR}, $z_{\mathrm{IR}}$ retains a dependence on the chemical potential even in the zero-temperature limit. We will return to this point in the following subsection.

\begin{figure}[h]
\begin{center}
\includegraphics[width=6.8cm]{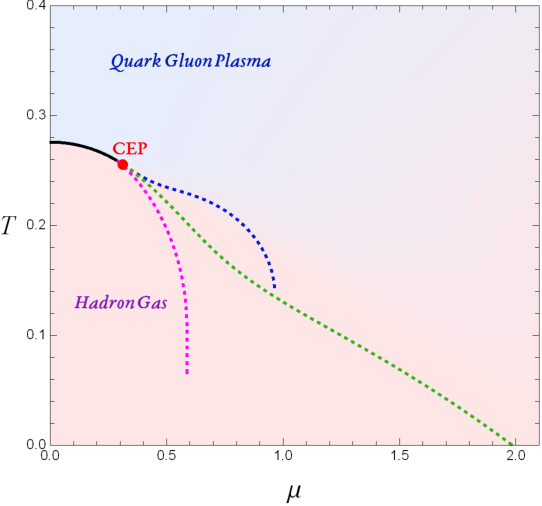}\hspace{.3cm}
\includegraphics[width=6.8cm]{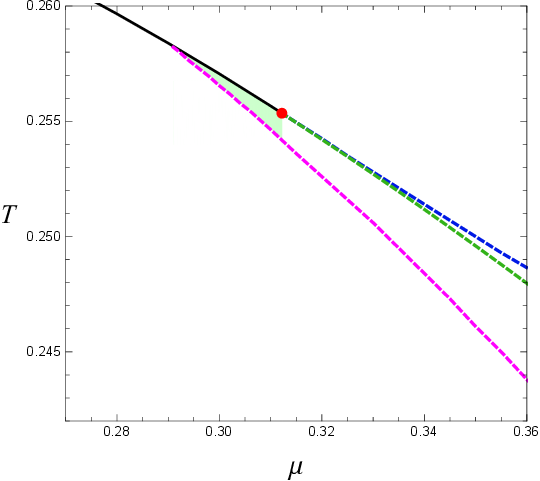}
\end{center}
\caption{\footnotesize 
Revised $T-\mu$ phase diagram of the model with logarithmic scale factor. The separation lines given by the presence of the IR wall is depicted by the magenta line. Right panel shows the enlarged region close to the CEP.}
\label{phasediagram2}
\end{figure}

As shown in the left panel of Fig.\,\ref{IR}, the IR wall persists at $\mu=0.4\,\mathrm{GeV}$, even though this value is larger than the critical chemical potential. The persistence of the IR wall can therefore serve as a dynamical signature for identifying crossover lines between confined-like and deconfined-like phases in the supercritical region. Such dynamical crossovers, which are not accompanied by any thermodynamic features, are similar to what commonly referred to as Frenkel lines \cite{ref11baggioli} and provide additional information about the structure of the phase diagram. The Frenkel-like line is indicated by the magenta dotted curve in the revised phase diagram of the second model in Fig.\,\ref{phasediagram2}, and it disappears abruptly at $\mu_c^{{\mathrm{IR}}}\approx 0.587\,\mathrm{GeV}$. Interestingly, as highlighted in the right panel showing the vicinity of the CEP, this line does not emanate directly from the critical end point. Instead, it separates from the first-order transition line at $(\mu,T)\approx(0.291\, \mathrm{GeV},0.258\,\mathrm{GeV})$. This behavior mirrors that of classical fluids, where dynamical crossover lines do not terminate at the critical point \cite{superholoBaggioli,stemnotcep}. Consequently, a region appears in the phase diagram below the CEP (the light green triangular area), which is confined thermodynamically but deconfined dynamically. 

\subsection{Critical electric fields}

In the presence of an external electric field $E$, the dynamics of the holographic system are governed by two distinct critical field strengths: $E_s$ and $E_c$. These fields probe different physical mechanisms and are determined through complementary holographic constructions.  

The first quantity, $E_s$, is defined as the minimal electric field strength required to overcome the confining force between a massless static quark–antiquark pair in the confined phase. In the holographic picture, $E_s$ is directly related to the position of the IR wall $z_{\mathrm{IR}}$, which represents the deepest radial point accessible to a connected U-shaped string. The effective string tension at $z_{\mathrm{IR}}$ sets the energy scale for breaking the confining flux tube, leading to  
\begin{equation}
   E_s \;=\; T_f \, \sigma_s(z_{\mathrm{IR}}),
   \label{eq:Es}
\end{equation}
where $\sigma_s(z)$ is the effective string tension density introduced in the previous subsection. Consequently, $E_s$ inherits the dependence of $z_{\mathrm{IR}}$ on temperature and chemical potential. In the first model, where $z_{\mathrm{IR}}$ is fixed, $E_s$ remains constant, while in the second model $E_s$ varies nontrivially with these parameters. In the deconfined phase, the IR wall disappears and thus $E_s=0$.  

Physically, $E_s$ is the largest electric field strength at which massive quarks face with an infinitly large total potential barrier, i.e., at $E=E_s$ the total potential of massive quarks becomes flat at infinity. Thus, the value of $E_s$ can also be found using $\lim_{L\to \infty} \frac{d V_{\mathrm{total}}}{d L}= 0$.
Therefore, $E_s$ can be viewed as the critical field at which the confining force is exactly canceled, although actual pair creation of massive quarks still requires stronger fields. 

The second critical electric field, $E_c$, characterizes the onset of catastrophic vacuum instability due to pair creation via the holographic Schwinger effect. It can be evaluated by imposing the reality condition of the Dirac–Born–Infeld (DBI) action of a probe D3 brane located at $z=z_0$, which encodes quarks of finite mass. With an external electric field applied along one spatial boundary direction, the DBI action reads  
\begin{align}
   S_{\mathrm{DBI}} &= - T_{D3} \int d^4x \, e^{-\phi(z_0)}
   \sqrt{-\det\!\left[ (g_s)_{\mu\nu}(z_0) + 2\pi\alpha' F_{\mu\nu} \right]}\nonumber\\
   &= - T_{D3} \int d^4x \, e^{-\phi(z_0)} {\cal H}_s(z_0)\sqrt{G(z_0){\cal H}_s(z_0)^2-(2\pi \alpha' E)^2}
   \label{eq:DBI_action}
\end{align}
where $T_{D3}$ is the D3-brane tension, $(g_s)_{\mu\nu}$ is the string-frame induced metric on the brane worldvolume, $\phi$ is the dilaton, and $F_{\mu\nu}$ encodes the electric field. The action remains real only if  
\begin{equation}
   E \;\leq\; E_c, \qquad
   E_c \;=\; T_f \,\sigma_s(z_0).
   \label{eq:Ec}
\end{equation}
This condition follows from requiring that the square root in \eqref{eq:DBI_action} remains non-negative. As a result, $E_c$ depends only on the background geometry at the probe brane position $z_0$. Physically, $E_c$ corresponds to the least electric field strength at which the potential barrier for massive quarks completely vanishes.  Thus, it is equal to the field strength which satisfies $\lim_{L\to 0} \frac{d V_{\mathrm{total}}}{d L}= 0$.

In the confined phase one always finds $E_s < E_c$, so that the onset of instability for massless quarks precedes that of massive quarks. In the deconfined phase, however, the IR wall is absent and $E_s=0$, while $E_c$ remains finite. The contrasting temperature and chemical potential dependence of $E_s$ and $E_c$ in our two EMD models therefore provides a sensitive probe of how confinement is encoded in their respective bulk geometries.  

In the following, we present the numerical results for $E_s$ and $E_c$ in the two models, comparing their behavior across the confined and deconfined phases and highlighting the role of the IR wall structure discussed above.

\begin{figure}[h]
\begin{center}
\includegraphics[width=6.8cm]{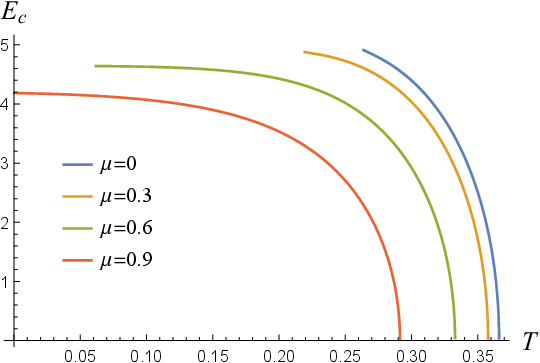}\hspace{.3cm}
\includegraphics[width=6.8cm]{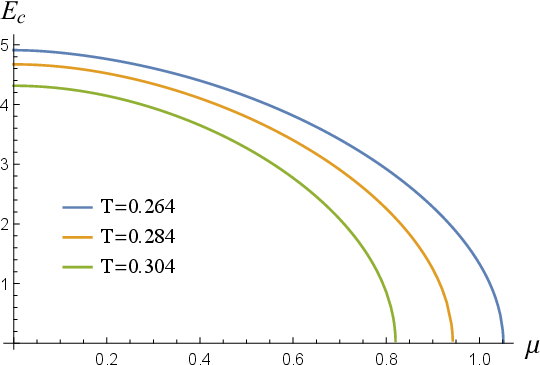}\hspace{.3cm}
\end{center}
\caption{\footnotesize 
Critical electric field $E_c$ for the deconfined phase of the first model with quadratic scale factor.}
\label{Ecq}
\end{figure}

The confined phase in the first model is described holographically by thermal AdS.
Since this background contains no black hole, all physical quantities are independent of temperature and chemical potential.
In particular, the IR wall remains fixed at $z_{\mathrm{IR}}=1.857\,\mathrm{GeV}^{-1}$, leading through Eqs.\,\eqref{eq:Es} and \eqref{eq:Ec} to constant values of the critical electric fields,
\[
E_s = 4.096\,\mathrm{GeV}^2 , \qquad E_c = 4.978 \,\mathrm{GeV}^2 ,
\]
after setting $z_0=1\,\mathrm{GeV}^{-1}$ without loss of generality.
In the deconfined phase, where the IR wall disappears, one finds $E_s=0$, while $E_c$ acquires a nontrivial dependence on $T$ and $\mu$, as displayed in Fig.\,\ref{Ecq}.
Both temperature and chemical potential reduce the value of $E_c$, making quark–antiquark pair production possible under a smaller external field.

\begin{figure}[h]
\begin{center}
\includegraphics[width=6.8cm]{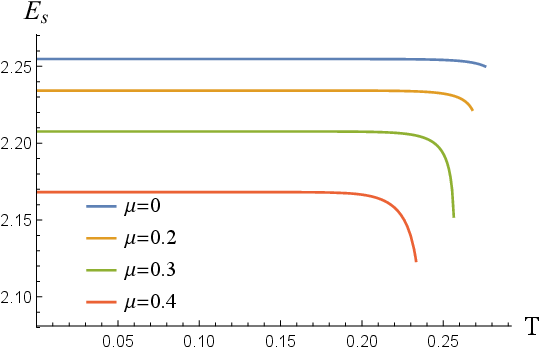}\hspace{.3cm}
\includegraphics[width=6.8cm]{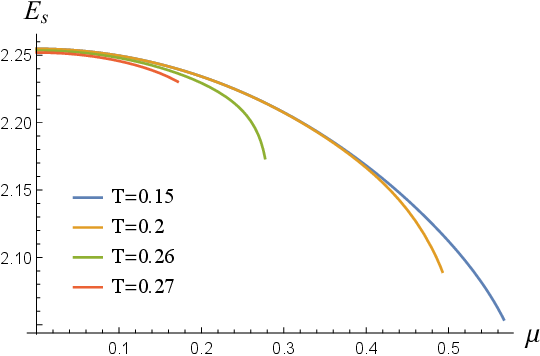}
\end{center}
\caption{\footnotesize 
Threshold electric field $E_s$ for the second model with logarithmic scale factor.}
\label{Eslog}
\end{figure}

In the second model, both confined and deconfined phases correspond to black hole geometries on the gravity side.
Therefore, even in the confined phase, quantities such as the IR wall position and the critical electric fields vary with temperature and chemical potential.
Figure \ref{Eslog} shows $E_s$ as a function of temperature (left) and chemical potential (right).
As expected from its relation to $z_{\mathrm{IR}}$, $E_s$ is nearly constant at low $T$ or $\mu$, but begins to decrease more noticeably as the system approaches the deconfinement transition.
At the transition, the IR wall disappears abruptly and $E_s$ drops discontinuously to zero.
Interestingly, a residual dependence on chemical potential persists even at $T\to 0$.
This deviation of $E_s$ (and $z_{\mathrm{IR}}$) from their zero-density values reflects the influence of finite charge density deep in the infrared.
Physically, this arises from explicit Lorentz symmetry breaking due to the presence of a finite density, which holographically corresponds to the nontrivial temporal component $A_t(z)$ of the bulk gauge field.

As is evident from Fig.\,\ref{Eslog}, the threshold electric field remains nonzero at low temperatures for $\mu=0.4\,\mathrm{GeV}$ in the supercritical region. This behavior is consistent with the results of the IR wall analysis presented in the previous subsection. Interestingly, throughout the entire region below the magenta line in Fig.\,\ref{phasediagram2}, we find $E_s \neq 0$, implying that the system responds to an external electric field in the same manner as in the confining phase. From the perspective of the Schwinger effect, this region of the phase diagram should therefore be regarded as confined. Notably, in the light green triangular area, although the system is thermodynamically confined, we obtain $E_s=0$. As a consequence, any arbitrarily small external electric field induces the liberation of massless quarks, resembling the behavior of the deconfined phase. 

\begin{figure}[h]
\begin{center}
\includegraphics[width=6.8cm]{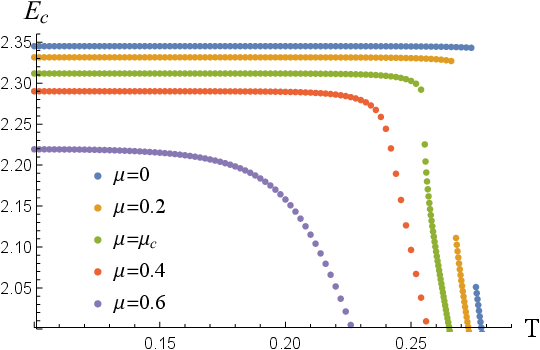}\hspace{.3cm}
\includegraphics[width=6.8cm]{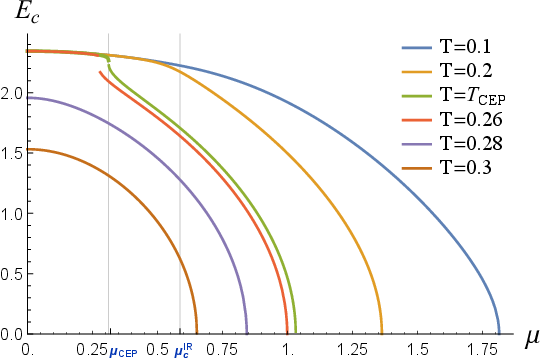}\hspace{.3cm}
\end{center}
\caption{\footnotesize 
Dependence of the Schwinger critical field $E_c$ on chemical potential and temperature for the second model with logarithmic scale factor. Across the first-order deconfinement transition $E_c$ exhibits a discontinuous drop, while a slope change appears only at $\mu_c^{IR}$ where the IR wall disappears. In confined and confined-like regimes $E_c$ depends only weakly on $T$ and $\mu$, whereas in the deconfined phase it decreases rapidly, making it a sensitive probe of the underlying phase structure.
}
\label{Eclog}
\end{figure}

The dependence of the Schwinger critical field $E_c$ on temperature and chemical potential in the second model is shown in Fig.\,\ref{Eclog}. 
In contrast to the threshold field $E_s$, which is obtained from the effective string tension at the IR wall position $z_{\mathrm{IR}}$ and vanishes once $z_{\mathrm{IR}}$ disappears in the deconfined phase, $E_c$ is determined at the probe brane position $z_0$ (fixed here to set the quark mass) and thus does not couple directly to the IR wall. Nevertheless, its behavior reflects the underlying phase structure.

Across the first-order deconfinement line, $E_c$ exhibits a discontinuous decrease, consistent with the nature of the phase transition. For instance, at $T=0.26\,\mathrm{GeV}$ the transition occurs at $\mu\simeq0.277\,\mathrm{GeV}$, where the discontinuity is clearly visible in the right graph. At the critical endpoint $\mu_{ _{\mathrm{CEP}}}$, indicated by the vertical line, $E_c(\mu)$ for the system with $T=T_{ _{\mathrm{CEP}}}$ shows a sudden drop. Beyond the CEP, in the supercritical region, for $\mu_{ _\mathrm{CEP}}<\mu<\mu_c^{\mathrm{IR}}$ with $\mu_c^{\mathrm{IR}}\approx0.587\,\mathrm{GeV}$, the system remains in a confined-like phase at sufficiently low temperatures, such as $T=0.1\,\mathrm{GeV}$ and $0.2\,\mathrm{GeV}$. 
In this regime, $E_c$ shows only mild variation with $\mu$, because the persistence of the IR wall maintains an approximately fixed string tension at $z_0$. 
Interestingly, a noticeable slope change arises upon crossing the dynamical separation line defined by the persistence of the IR wall, where the IR wall itself disappears; although $E_c$ is not directly defined at $z_{\mathrm{IR}}$, the loss of the wall modifies the background geometry and indirectly influences the effective string tension at $z_0$.  

At higher temperatures, such as $T=0.28\,\mathrm{GeV}$ and $0.30\,\mathrm{GeV}$, the system is fully deconfined even at small chemical potential, and $E_c$ decreases more steeply with both $T$ and $\mu$ and without any distinct feature. This contrast highlights the role of the phases: in confined and confined-like regimes, $E_c$ is almost insensitive to external parameters, consistent with the stability of the confined phase against pair production, whereas in the deconfined phase it responds strongly to changes in $T$ and $\mu$, showing that the medium facilitates quark--antiquark creation in the deconfined regime. Thus, although less directly tied to $z_{IR}$ than $E_s$, the critical field $E_c$ remains a valuable diagnostic of the QCD phase diagram, carrying clear imprints of first-order transitions, the CEP, and the IR-wall--driven crossover line in the supercritical region.

\subsection{Total Potential}

We now explore the response of our holographic QCD models to an external electric field by analyzing the total potential of a $q\bar{q}$ pair in the presence of $E$.

\begin{figure}[h]
\begin{center}
\includegraphics[width=6.8cm]{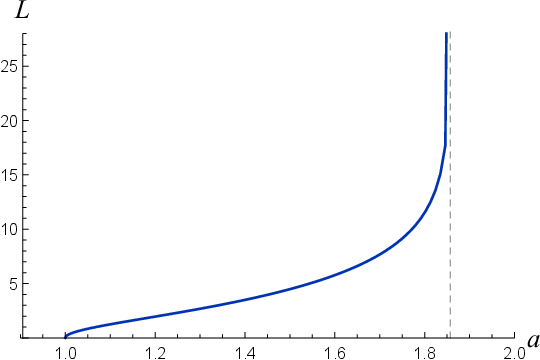}\hspace{.3cm}
\includegraphics[width=6.8cm]{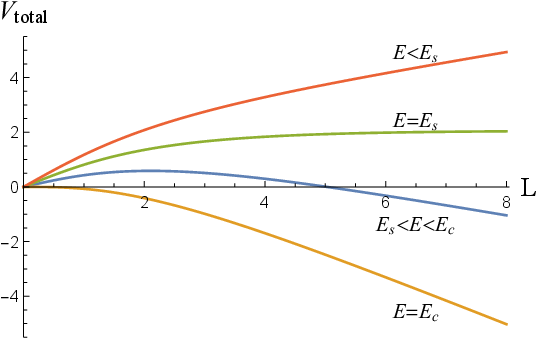}
\end{center}
\caption{\footnotesize 
The separation length of the quarks versus the rescaled radial position of the turning point of the string (left) and the total potential versus the separation length (right) in the confined phase of the first model with quadratic scale factor.}
\label{LVconfq}
\end{figure} 

We begin with the confined phase of the first model. The left panel of Fig.\,\ref{LVconfq} shows the separation length between the quark and antiquark, $L$, as a function of $a$, the rescaled position of the tip of the hanging string with respect to $z_0$. As discussed earlier, all physical quantities in this phase are independent of temperature and chemical potential. Increasing $a$ increases the quark--antiquark separation. In the confined phase, the connected U-shaped configuration is always the preferred solution. Thus, as the separation grows, the string never breaks; instead, its tip penetrates deeper into the bulk until it reaches the IR wall, where it lies flat. In the separation-length graph, this appears as an asymptotic divergence of $L$ at the IR wall position, which signals the linear rise of the $q\bar{q}$ potential at large distances—characteristic of confinement.

The right panel of Fig.\,\ref{LVconfq} displays the total potential as a function of $L$ for various values of the electric field. For $E<E_s$, the potential barrier is infinite and no pairs can be produced. At $E=E_s$, the potential barrier becomes flat at infinite separation, defining the threshold for pair production. Increasing $E$ further reduces the barrier: for $E_s<E<E_c$, pair creation can occur via quantum tunneling. At $E=E_c$, the potential barrier for quarks of fixed mass vanishes, and vacuum decay sets in. For $E>E_c$, the system becomes unstable, and quarks of unit and lower masses are produced without suppression.

\begin{figure}[h]
\begin{center}
\includegraphics[width=6.8cm]{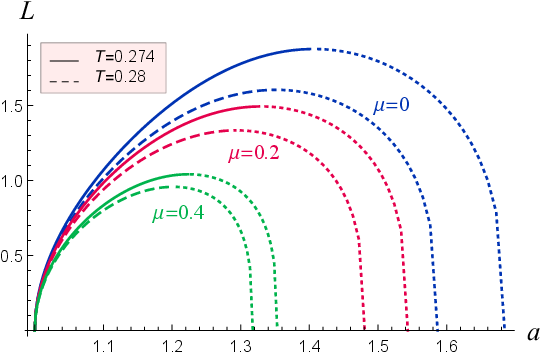}\hspace{.3cm}
\includegraphics[width=6.8cm]{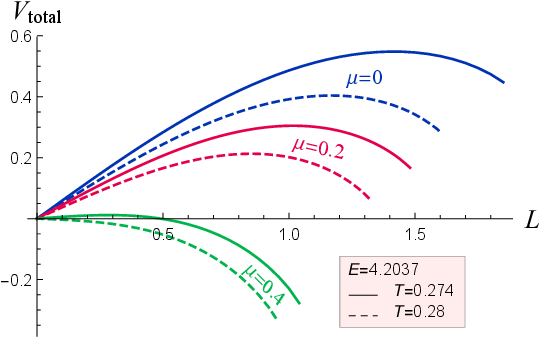}
\end{center}
\caption{\footnotesize 
The separation length of the quarks versus the rescaled radial position of the turning point of the string (left) and the total potential versus the separation length (right) in the deconfined phase of the first model with quadratic scale factor, for various values of temperature and chemical potential.}
\label{LVdeconfq}
\end{figure} 

Next, consider the deconfined phase of the first model. Here, $E_s=0$, so even an arbitrarily small electric field can induce the creation of massless quarks. On the gravity side, this phase corresponds to a charged black hole solution, and physical observables depend on both $T$ and $\mu$. Figure \ref{LVdeconfq} shows $L(a)$ and $V_{\mathrm{total}}(L)$ for various chemical potentials at two different temperatures. In contrast to the confined case, there exist two connected U-shaped string solutions for each separation length. The energetically favorable branch corresponds to the configuration whose turning point lies closer to the D3 brane. (These are displayed by solid and dashed curves and the other solutions are plotted as dotted curves.) Notably, there is a maximum interquark distance $L_{\mathrm{max}}$, beyond which the connected configuration ceases to exist, signaling dissociation of the $q\bar{q}$ pair into free quarks (disconnected string configuration). This absence of linear confinement is reflected in both $L(a)$ and the total potential (drawn here for $E=4.2037\,\mathrm{GeV}^2$). Increasing temperature or chemical potential reduces $L_{\mathrm{max}}$ and lowers both the height and width of the potential barrier, weakening the binding between quarks.

\begin{figure}[h]
\begin{center}
\includegraphics[width=6.8cm]{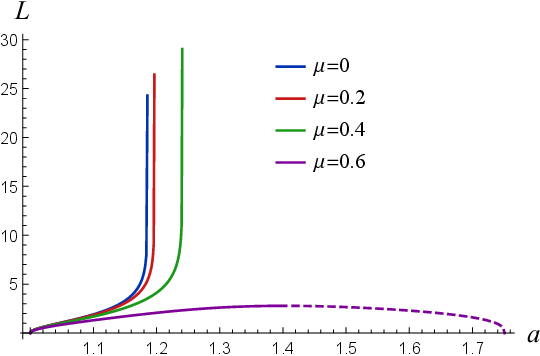}\hspace{.3cm}
\includegraphics[width=6.8cm]{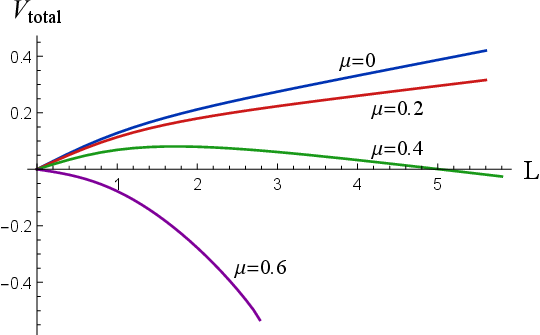}
\end{center}
\caption{\footnotesize 
$L$ versus the rescaled radial position $a$ (left) and the total potential versus $L$ (right) for different values of $\mu$ in the second model with logarithmic scale factor. In all cases $T=0.2\,\mathrm{GeV}$ and the applied electric field in the right panel is $E=2.2\,\mathrm{GeV}^2$.}
\label{LVlog1}
\end{figure} 

We now turn to the second model, which admits two phases: the specious confined phase (dual to a small black hole) and the deconfined phase (dual to a large black hole). In both, the response to an external electric field depends on $T$ and $\mu$. The deconfined phase behaves similarly to that of the first model, qualitatively, but the specious confined phase displays distinct features. 

Figure \ref{LVlog1} shows $L(a)$ and $V_{\mathrm{total}}(L)$ for $T=0.2\,\mathrm{GeV}$ and various chemical potentials, with the total potential computed at $E=2.2\,\mathrm{GeV}^2$. Increasing $\mu$ drives the system closer to deconfinement, easing the transition to the deconfined phase. For $\mu<\mu_{ _{\mathrm{CEP}}}$, $L(a)$ behaves like in the confined phase. Interestingly, even at $\mu=0.4\,\mathrm{GeV}$, within the supercritical region, the system still responds as if confined—consistent with the IR wall separation line discussed earlier. From the revised phase diagram in Fig.\,\ref{phasediagram2}, we expect that for $\mu>0.587175\,\mathrm{GeV}$ (beyond the endpoint of the IR wall line), the system behaves like the deconfined phase in its response to electric fields. This is evident in the left panel of Fig.\,\ref{LVlog1} for $\mu=0.6\,\mathrm{GeV}$. In all cases—confined, confined-like (supercritical), and deconfined—increasing $\mu$ reduces the barrier height and width, making quark liberation easier.

\begin{figure}[h]
\begin{center}
\includegraphics[width=6.8cm]{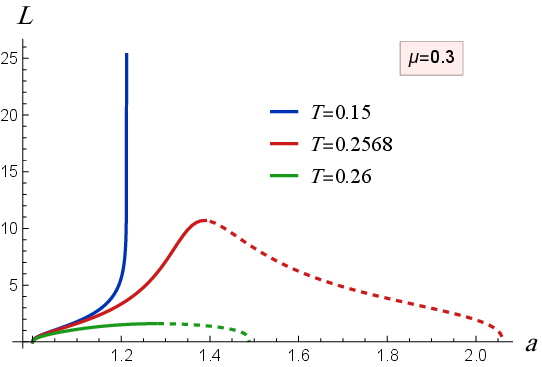}\hspace{.3cm}
\includegraphics[width=6.8cm]{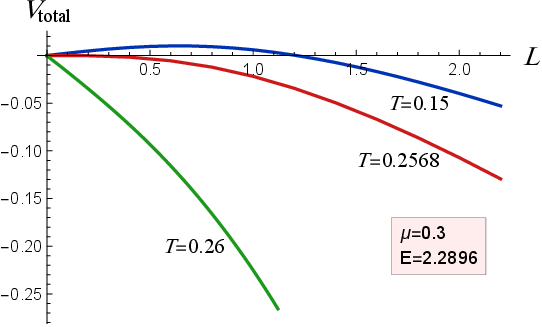}
\end{center}
\caption{\footnotesize 
$L(a)$ (left) and $V_{\mathrm{total}}(L)$ (right) at $\mu=0.3\,\mathrm{GeV}$ for different values of temperature in the second model with logarithmic scale factor. The applied electric field in the right panel is $E=2.2896\,\mathrm{GeV}^2$.}
\label{LVlog2}
\end{figure}

Finally, to probe the triangle region of the phase diagram, Fig.\,\ref{LVlog2} displays $L(a)$ for $\mu=0.3\,\mathrm{GeV}$ and $T=0.1\,\mathrm{GeV}$ (thermodynamically and dynamically confined), $T=0.2568\,\mathrm{GeV}$ (inside the triangle), and $T=0.2\,\mathrm{GeV}$ (above both the first-order transition line and the IR wall line). In the right panel of Fig.\,\ref{LVlog2}, the corresponding total potentials are shown at $E\simeq 2.2896\,\mathrm{GeV}^2$, equal to $E_c$ for the second case. Inside the triangle region, $L(a)$ resembles the deconfined phase but with a very large $L_{\mathrm{max}}$. As in the other cases, increasing $T$ lowers the potential barrier, allowing easier quark liberation. 

To summarize, the analysis of the total potential provides a complementary dynamical perspective on the phase structure of the models. In the confined phases, the infinite potential barrier and non-vanishing $E_s$ signal the stability of bound states, while in the deconfined phase, the disappearance of $E_s$ and the existence of a finite $L_{\mathrm{max}}$ demonstrate screening and pair dissociation. The specious confined phase extends these behaviors into the supercritical region, where the system can still respond as if confined. Taken together with the thermodynamic probes (specific heat) and dynamical ones (speed of sound, IR wall), the total potential analysis reinforces the existence of distinct crossover lines beyond the CEP, highlighting the richness of the holographic phase diagram and the utility of the Schwinger effect as a sensitive diagnostic tool. 

\section{Summary and conclusion}

When dealing with strongly coupled non-perturbative systems where perturbative field-theoretical methods face limitations, or in the presence of finite baryon density where conventional approaches such as lattice QCD are unreliable, holography offers a powerful alternative without such restrictions. 

In this work, we employed a specific Einstein--Maxwell--Dilaton (EMD) model proposed by Dudal and Mahapatra, which constructs black hole solutions by tuning the scale factor so as to reproduce expected properties on the dual field theory side. They introduced quadratic and logarithmic scale factors, giving rise to standard confinement/deconfinement and specious confinement/deconfinement transitions, respectively. A key strength of the specious confined phase lies in its capacity to explore the temperature and density dependence of physical quantities in the confined phase—a regime largely inaccessible to other theoretical frameworks. Furthermore, as Dudal and Mahapatra showed, the phase diagram of the second model features a first-order transition from the specious confined phase (hadronic gas) to the deconfined phase (quark--gluon plasma), which terminates at a second-order critical end point (CEP). This structure provides a rich phase diagram, including a supercritical region whose properties remain largely unexplored, and raises the question of whether separation lines of thermodynamic or dynamical origin can be used to distinguish phases beyond the critical point.

The primary objective of the present study was to further unravel the QCD phase structure and the properties of QCD matter using these holographic models. After briefly reviewing the thermodynamics of the second model (with the first model as a comparison), we analyzed the heat capacity and speed of sound in the second model to track the transition from confined to deconfined phases. The heat capacity exhibits a peak at the deconfinement transition temperature: smooth and of small magnitude in the distant supercritical region, but sharper and larger near the CEP, where it diverges. Below the CEP, it shows delta-function-like behavior at the first-order transition temperature. By contrast, the squared sound speed displays a shallow minimum far from the CEP, which deepens and approaches zero near the CEP; at the CEP itself, it vanishes. Thus, the lowest minimum of $C_s^2$ is precisely at the second-order critical point. The loci of extrema of these two observables for $T>T_{ _{\mathrm{CEP}}}$ introduce two distinct crossover separation lines, both emanating from the CEP and extending into the supercritical region. However, they trace different paths: the heat-capacity line lies at higher temperatures and drops sharply to zero at finite $\mu$, while the sound-speed line smoothly continues to a higher $\mu$, intersecting the density axis.

We also identified another dynamical separation line via the IR wall in the bulk, which represents the deepest radial point accessible to a hanging string. The presence of an IR wall is a hallmark of confinement. Remarkably, our analysis shows that the IR wall persists even beyond the CEP. The corresponding separation line does not originate from the CEP, in contrast to the thermodynamic probes and the speed of sound, but is consistent with analogous findings for classical fluids. Moreover, a small triangular region appears below the CEP where the system is confined thermodynamically but not dynamically according to the IR wall.

The Schwinger effect was also investigated in both models. On the one hand, it provides a sensitive dynamical probe capable of distinguishing confined and deconfined phases through critical electric fields, offering complementary insight into confinement and phase transitions of QCD matter. On the other hand, the rich phase structure of the second model allows us, for the first time in holography (to the best of our knowledge), to explore the thermal and chemical potential dependence of the Schwinger effect in confined, deconfined, and supercritical regimes.

We determined both the critical electric field $E_c$, above which quarks of a given mass can be freely produced, and the threshold electric field $E_s$, below which the Schwinger effect is absent, in both models. In the specious confined phase, unlike the standard confined phase, these fields depend nontrivially on $T$ and $\mu$. Both temperature and chemical potential decrease $E_c$ and $E_s$, thereby facilitating the Schwinger effect. In particular, $E_s$ closely tracks the IR wall position: nearly constant at low $T$, but decreasing more significantly near $T_c$ (either the first-order transition temperature or the transition temperature given by the IR wall in the supercritical region), where it remains finite and nonzero; just above $T_c$, it discontinuously jumps to zero. We also observed a nontrivial dependence of $E_s$ and $z_{\mathrm{IR}}$ on chemical potential in the $T\to 0$ limit, signaling that finite density influences persist deeply in the infrared, even in the absence of thermal effects. This can be naturally understood as a manifestation of Lorentz symmetry breaking induced by finite charge density. Furthermore, we were able to trace $E_s$ in the supercritical region, finding that the system responds as confining to external electric fields for $(T,\mu)$ values below the IR-wall separation line.

In addition to the threshold field $E_s$, the critical field $E_c$ also carries clear signatures of the QCD phase structure. While less directly tied to the IR wall, $E_c$ shows sharp drops across first-order transitions, the CEP, and a slope change at the IR-wall separation line in the supercritical region. Its weak dependence in confined regimes and strong sensitivity in the deconfined phase also underline its diagnostic value: $E_c$ provides complementary evidence, alongside $E_s$, for identifying phase boundaries and critical phenomena in the QCD diagram. 

Finally, we analyzed the quark--antiquark separation length and the total potential in various regimes. In all cases, increasing $T$ or $\mu$ decreases both the height and width of the potential barrier, enhancing pair production. A particularly notable result arises in the triangular region of the phase diagram: although the system is thermodynamically confined, the interquark distance behaves as in the deconfined phase, with very large $L_{\mathrm{max}}$, highlighting the subtle interplay between thermodynamic and dynamical confinement.


Given the importance of supercritical crossover studies in the QCD phase diagram, it would be valuable to extend holographic investigations using other QCD-like models, particularly those with flavor degrees of freedom, to test the universality of our findings. Moreover, additional probes beyond thermodynamic quantities and the Schwinger effect could provide complementary diagnostics. In particular, entanglement entropy and mutual information have already proven to be sensitive indicators of confinement and could reveal new insights into the structure of the supercritical region. Exploring such probes, together with transport coefficients and non-equilibrium observables, may help to establish a more complete picture of QCD matter beyond the critical point and clarify the interplay between thermodynamic and dynamical signatures of confinement.


 \end{document}